\documentstyle[multicol,twoside,eqsecnum,epsf,aps,prb]{revtex}

\begin{document}
\newcommand{\fig}[2]{\epsfxsize=#1\epsfbox{#2}} \reversemarginpar 

\newenvironment{tab}[1]
{\begin{tabular}{|#1|}\hline}
{\hline\end{tabular}}

\title{Strongly disordered spin ladders}

\author{R. M\'elin$^{1}$,Y.-C. Lin$^{2}$, P. Lajk\'o$^{3}$,
H. Rieger$^{4}$, and F. Igl\'oi$^{1,3,5}$}

\address{
$^1$ Centre de Recherches sur les Tr\'es Basses
Temp\'eratures\thanks{U.P.R. 5001 du CNRS, Laboratoire
conventionn\'e avec l'Universit\'e Joseph Fourier},
B. P. 166,
F-38042 Grenoble, France\\
$^2$ NIC, Forschungszentrum J\"ulich, 52425 J\"ulich, Germany\\
$^3$ Institute for Theoretical Physics,
Szeged University, H-6720 Szeged, Hungary\\
$^4$ Theoretische Physik, Universit\"at des Saarlandes, 
     66041 Saarbr\"ucken, Germany\\
$^5$ Research Institute for Solid State Physics and Optics, 
H-1525 Budapest, P.O.Box 49, Hungary\\
}

\date{\today}

\maketitle

\begin{abstract}
The effect of quenched disorder on the low-energy properties of various
antiferromagnetic spin ladder models is studied by a numerical strong disorder
renormalization group method and by density matrix renormalization.
For strong enough disorder
the originally gapped phases with finite topological or dimer order become
gapless. In these quantum Griffiths phases the scaling of the energy, as well as
the singularities in the dynamical quantities are
characterized by a finite dynamical exponent, z, which varies with the strength
of disorder. At the phase boundaries, separating topologically distinct
Griffiths phases the singular behavior of the disordered ladders is generally
controlled by an infinite randomness fixed point.
\end{abstract}


\newcommand{\bc}{\begin{center}}
\newcommand{\ec}{\end{center}}
\newcommand{\be}{\begin{equation}}
\newcommand{\ee}{\end{equation}}
\newcommand{\ba}{\begin{array}}
\newcommand{\ea}{\end{array}}
\newcommand{\beqn}{\begin{eqnarray}}
\newcommand{\eeqn}{\end{eqnarray}}
\begin{multicols}{2}
\narrowtext

\section{Introduction}
Low dimensional quantum spin systems, chains and ladders are fascinating objects,
which are the subject of intensive experimental and theoretical research. The main
source of this activity is due to the observation that quantum fluctuations could
result in qualitatively different low-energy behavior in these interacting
many-body systems. It was first Haldane\cite{haldane} who conjectured that
antiferromagnetic (AF) spin chains with integer spin have a gap in the energy
spectrum (Haldane phase), whereas the spectrum of chains with half-integer
spins is gapless. By now a large amount of experimental and theoretical
evidence have been collected in favor of the Haldane conjecture. It has
been realized by Affleck, Kennedy, Lieb and Tasaki\cite{AKLT} (AKLT) that the ground
state structure of the Haldane phase for $S=1$ is closely related to that of the
valence-bond solid model, where the ground state is built up from nearest-neighbor
valence bonds. The hidden, topological order in the chain is measured by the
non-local string order parameter\cite{nijs}:
\be
O^{\alpha}=-\lim_{|i-j| \to \infty} \left\langle S_i^{\alpha} \exp\left(i \pi
\sum_{l=i+1}^{j-1} S_l^{\alpha} \right) S_j^{\alpha} \right\rangle\;,
\label{string1}
\ee
where $S_i^{\alpha}$ is a spin-1 operator at site $i$, $\alpha=x,y,z$
and $\langle \dots \rangle$ denotes the ground state expectation value.

Another source of activity in the field of low-dimensional quantum spin systems
is due to the discovery of spin
ladder materials\cite{rice}. It has been realized that spin ladders with
even number of legs have a gapped spectrum, whereas the spectrum of
odd-leg ladders is gapless\cite{scalapino92}. For two-leg ladders, which
are analogous objects to $S=1$ spin chains, the ground state structure
can be related to nearest-neighbor valence bonds and a topological hidden
order parameter, similar to that in Eq.(\ref{string1}) can be
defined\cite{kim}. 

More recently, ladder models with competing interactions,
such as with staggered
dimerization\cite{delgado} and with rung and diagonal couplings\cite{kim},
have been introduced and studied. In these models, depending on the relative
strength of the couplings, there are several gapped phases with
different topological   
order, which are separated by first- or second-order phase transition lines.

Disorder turns out to play a crucial role in some experiments
on low dimensional magnets. For instance,
the NMP-TCNQ compound\cite{TCNQ} can be well described by
$S=1/2$ spin chains with
random AF couplings. More recently, non magnetic substitutions
in low dimensional oxydes such as
CuGeO$_3$\cite{CuGeO3-1,CuGeO3-2,CuGeO3-3,CuGeO3-4}
(being a spin-Peierls compound),
PbNi$_2$V$_2$O$_8$\cite{Pb} (being a Haldane gap compound)
or Y$_2$BaNiO$_5$\cite{ori-Y,Batlogg,Y2-1,Y2-2,Y2-3,Y2-4}
(being a Haldane gap compound)
have been the subject of intense investigations. The essential
feature of these compounds is the appearance of
antiferromagnetism at low temperature which can be well
described by the effective low energy models introduced
in Refs.\onlinecite{Y2-4,Fabrizio-Melin,Melin-RG}.
Sr(Cu$_{1-x}$Zn$_x$)$_2$ O$_3$ 
is a realization of the two-leg ladder, and can be doped
by Zn, a non magnetic ion\cite{azuma}.
The specific heat and spin
susceptibility experiments indicate that the doped system
is gapless even with low doping concentrations.
We note that the experimentally found phase diagram of this compound,
as well as other
quantities, such as staggered susceptibility have been obtained by quantum
Monte Carlo simulations\cite{miyazaki}.

Theoretically, spin chains in the presence of strong disorder
can be conveniently studied by a real-space renormalization group (RG) method
introduced by Ma, Dasgupta and Hu\cite{mdh} (MDH). In this method strong bonds
in the system are successively eliminated and other bonds are replaced by weaker
ones through a second order perturbation calculation. As realized later by
Fisher\cite{fisherxx} for the random spin-1/2 chain and for the related model of
random transverse-field Ising spin chain\cite{fisher} the probability distribution
of the couplings under renormalization becomes broader and broader without limit
and therefore the
system scales into an infinite randomness fixed point (IRFP), where the MDH
renormalization becomes asymptotically exact. Fisher has also succeeded to
solve the fixed-point RG equations in analytical form and to show, that for
any type of (non-extremely singular) initial disorder the system scales into
the same IRFP. Later numerical\cite{girvin,bigpaper,ijr00} and analytical\cite{ijr00}
work has confirmed Fisher's results. 

Generalization of the MDH approach for AF chains with larger values of the spin is not
straightforward, since for not too strong initial disorder the generated
new couplings could exceed the value of the already decimated ones. To
handle this problem for the $S=1$ chain Hyman and Yang\cite{hyman} and independently
Monthus, Golinelli and Jolicaeur\cite{monthus} have introduced an
effective model with spin-1 and spin-1/2 degrees of freedom and with
random AF and ferromagnetic (F) couplings. From an analysis
of the RG equations they arrived to the conclusion that the IRFP of the
model will be attractive if the original distribution parametrized
by the power-law form
\be
P_{pow}(J)={1 \over D} J^{-1+1/D}\;.
\label{pow_dist}
\ee
is strongly random, i.e. if $0<D^{-1}<D_1^{-1}$. For weaker initial
disorder the system is still gapless, which is called the gapless
Haldane phase.

Theoretical work about disordered spin ladders is mainly concentrated
on the weak disorder limit. Results in this direction are obtained
in the weak interchain coupling limit via the bosonization
approach\cite{giamarchi} and by the random mass Dirac fermion
method\cite{gogolin}. In particular a remarkable stability of the
phases of the pure system against disorder with $XY$ symmetry
has been observed\cite{giamarchi}.

In the experimental situation, however, as described before the
effect of disorder is usually strong and we are going to
consider this limiting case in this paper. Our aim is to provide a
general theoretical background for strongly disordered spin ladders
by studying in detail several models (conventional ladder, dimerized
ladder, zig-zag ladder, and the full ladder with rung and diagonal
couplings), which could have experimental relevance. Since often a small
change in the couplings or in the strength of disorder
could cause large differences in the low-energy singular properties of the
models, therefore we have studied the phase diagrams in the space of several
parameters. As a method of calculation we used a
numerical implementation of the MDH approach, which could treat the
combined effect of disorder, frustration, correlations and quantum fluctuations,
while some problems are also studied by density matrix renormalization (DMRG).
In particular we have investigated the stability of the different topologically
ordered phases and studied the region of attraction of the IRFP.

The structure of the paper is the following. In Sec. 2 we define different
spin ladder models and present their phase diagram for non-random couplings.
A short overview about the MDH RG method and its application to random
spin chains are given in Sec. 3. Our results about random spin
ladders are presented in Sec. 4 and discussed in Sec. 5.

\section{The models and their phase diagram for non-random couplings}

We start with two spin-1/2 Heisenberg chains,
labeled by $\tau=1,2$ and
described by the Hamiltonians
\be
H_{\tau}=\sum_{l=1}^L J_{l,\tau} {\bf S}_{l,\tau} {\bf S}_{l+1,\tau}\;,
\label{Htau}
\ee
where ${\bf S}_{l,\tau}$ is a spin-1/2 operator at site $l$ and on chain $\tau$
and $J_{l,\tau}>0$. For non-random spin chains dimerization can be introduced
as
\be
J_{l,\tau}=J\left[1+\gamma (-1)^{l+n(\tau)}\right],\quad 0\le \gamma < 1\;,
\label{Jdimer}
\ee
with $n(\tau)=0,1$, whereas for random dimerized couplings the even and odd bonds
are taken from different distribution.
The pure chain without dimerization ($\gamma=0$) has a gapless
spectrum, and spin-spin correlations decay as a power for
large distance, which is called as quasi-long-range-order (QLRO).
Introducing dimerization for $\gamma>0$ a gap opens in the spectrum\cite{crossfisher}, which is
accompanied by non-vanishing dimer order, $O^{\alpha}_{dim}\ne 0$. This is measured as the
difference between the string order parameters in Eq(\ref{string1}) calculated
with spin-1/2 moments at even (e) and odd (o) sites:
\be
O^{\alpha}_{dim}=O^{\alpha}_{e}-O^{\alpha}_{o}\;.
\label{stringd}
\ee
In the following we generally consider non-dimerized chains, otherwise it is explicitly
mentioned.

\begin{figure}[h]
\epsfxsize=8truecm
\begin{center}
\mbox{\epsfbox{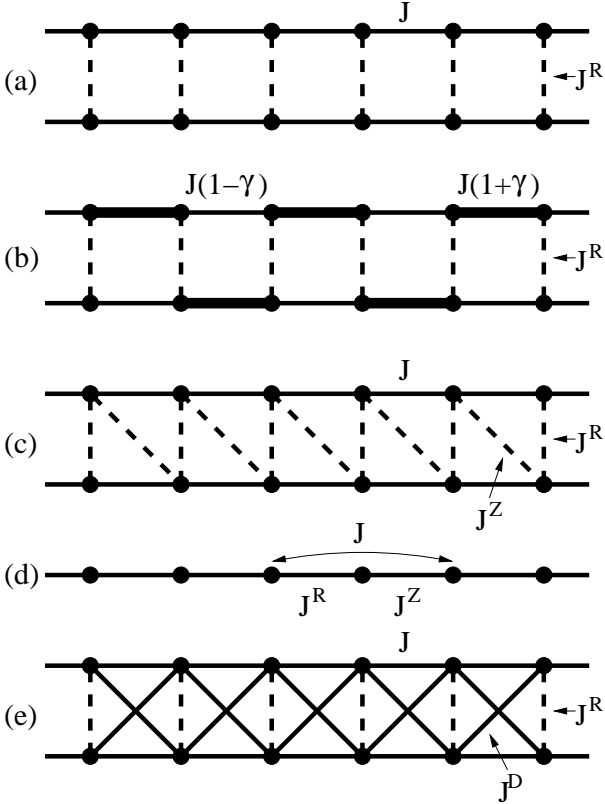}}
\end{center}
\caption{\label{fig1} Spin-ladder models used in the paper. The conventional
two-leg ladder (a) and 
staggered dimerization in the chain couplings (b). The zig-zag ladder (c) and its
representation as a chain with first and second neighbor couplings (d). The full ladder
with rung and diagonal couplings (e).}
\end{figure}

Now we introduce the interchain interaction
\be
H_{R}=\sum_{l=1}^L J_{l}^R {\bf S}_{l,1} {\bf S}_{l,2}\;,
\label{HR}
\ee
which describes the usual rung coupling between the ladders
(see Fig.~\ref{fig1}.a). The
conventional ladder model is described by the Hamiltonian: $H=H_1+H_2+H_R$.
In the pure model by switching on the AF rung couplings, $J_l^R=J^R>0$,
a Haldane-type gap opens above the ground state and the system has a non-vanishing
even string topological order, which is measured by\cite{kim,remark1}:
\end{multicols}
\widetext
\be
O^{\alpha}_{even}=-\lim_{|i-j| \to \infty} \left\langle (S_{i+1,1}^{\alpha}+ S_{i,2}^{\alpha})
\exp\left(i \pi \sum_{l=i+1}^{j-1} (S_{l+1,1}^{\alpha}+S_{l,2}^{\alpha})\right) 
(S_{j+1,1}^{\alpha} + S_{j,2}^{\alpha})\right\rangle\;.
\label{stringeven}
\ee
\begin{multicols}{2}
\narrowtext
For strong AF rung couplings every spin-pair on the same rung form
a singlet, therefore this phase is called the {\it rung singlet} (RUS) phase.

\begin{figure}[h]
\epsfxsize=8truecm
\begin{center}
\mbox{\epsfbox{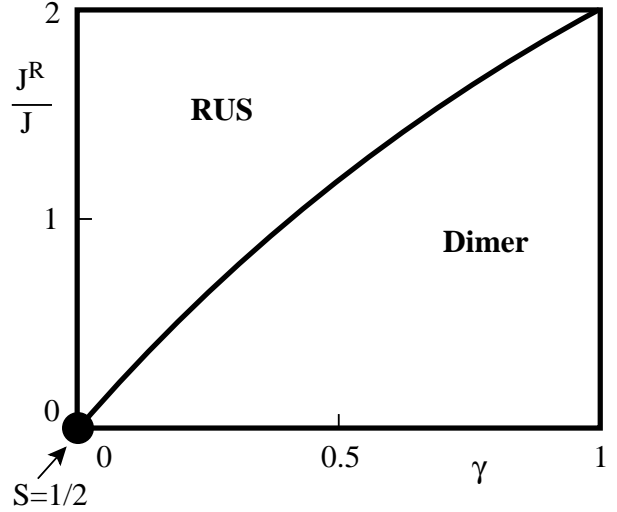}}
\end{center}
\caption{\label{fig2} Schematic phase diagram of the two-leg AF ladder with staggered
dimerization (see Fig.~\ref{fig1}.b for the definition of the couplings.) At the phase boundary
between the rung singlet and dimer phases the gap vanishes.}
\end{figure}

Dimerization of the chain couplings could
occur in two different ways. For
parallel dimerization, when equal bonds in the two
chains are on the same position, i.e. in Eq.(\ref{Jdimer}) $n(1)=n(2)$,
the combined effect of rung coupling and dimerization will always result in
a gapped phase. In the other possible case of staggered dimerization, i.e. with
$n(1)=-n(2)$ (see Fig.~\ref{fig1}.b), the two chains have an
opposite dimer order, which
competes with the rung coupling. As a result the phase diagram of the system
(see Fig.~\ref{fig2}) consists of two gapped phases, which are separated by a gapless
transition line, starting in the pure, decoupled chains limit\cite{delgado}. 

Next, we extend our model by diagonal interchain couplings, given by the Hamiltonian term:
\be
H_{Z}=\sum_{l=1}^L J_{l}^Z {\bf S}_{l,2} {\bf S}_{l+1,1}\;.
\label{HZ}
\ee
The complete Hamiltonian, $H=H_1+H_2+H_R+H_Z$, describes a zig-zag ladder (see Fig.~\ref{fig1}.c)
or can be considered as a spin chain with nearest neighbor ($J_l^R,~J_l^Z$) and next-nearest
neighbor ($J_l$) couplings (Fig.~\ref{fig1}.d). The pure model with $J_l^R=J_l^Z=J_1$
and $J_l=J_2$
has two phases: a gapless phase for $J_2/J_1<.24$ is separated from a gapped phase by a
quantum phase transition point.

\begin{figure}[h]
\epsfxsize=8truecm
\begin{center}
\mbox{\epsfbox{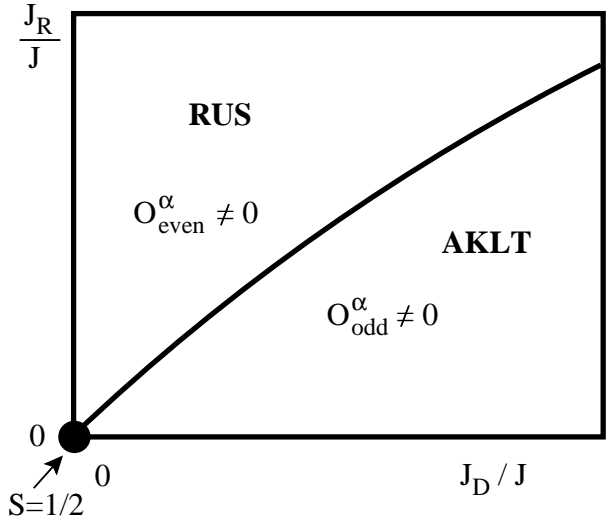}}
\end{center}
\caption{\label{fig3} Schematic phase diagram of the full AF ladder with homogeneous
rung and diagonal couplings. The transition between the two topologically distinct gapped
phases is of first order, except of the limit $J_R=J_D=0$.}
\end{figure}

Finally, we extend our model by two types of diagonal couplings, which are represented by
the Hamiltonian:
\be
H_{D}=\sum_{l=1}^L J_{l}^D ({\bf S}_{l,1} {\bf S}_{l+1,2} + {\bf S}_{l,2} {\bf S}_{l+1,1})\;.
\label{HD}
\ee
It is known that the pure AF diagonal ladder described by the Hamiltonian,
$H=H_1+H_2+H_D$ with $J_l^D=J^D>0$ has a gapped spectrum\cite{kim}. Its ground state is of the
AKLT-type and has a non-vanishing odd string order\cite{remark1}, defined in analogy to Eq.(\ref{stringeven})
\end{multicols}
\widetext
\be
O^{\alpha}_{odd}=-\lim_{|i-j| \to \infty} \left\langle (S_{i,1}^{\alpha} + S_{i,2}^{\alpha})
\exp\left(i \pi \sum_{l=i+1}^{j-1} (S_{l,1}^{\alpha}+S_{l,2}^{\alpha})\right) 
(S_{j,1}^{\alpha} + S_{j,2}^{\alpha})\right\rangle\;.
\label{stringodd}
\ee
\begin{multicols}{2}
\narrowtext
In the full ladder there are both rung and diagonal couplings
(see Fig.~\ref{fig1}.e) and it is described by
the Hamiltonian $H=H_1+H_2+H_R+H_D$. For non-random AF couplings there is
a competition between rung and diagonal couplings, so that the ground state
phase diagram of the system consists of two topologically distinct gapped phases
(see Fig.~\ref{fig3}). The
phase transition between the two phases is of first order\cite{kim}.

The main subject of our paper is to investigate how the phase diagrams of the pure ladder models,
in particular in
Figs.~\ref{fig2} and~\ref{fig3}
are modified due to the presence of quenched disorder.

\section{The MDH renormalization: results for spin chains}

In the MDH renormalization group method for random spin-1/2 chains the random AF
bonds are arranged in descending order according to their strength, and the strongest bond, say
$J_{23}$, connecting sites $2$ and $3$ sets the energy scale in the problem, $\Omega=J_{23}$.
We denote by $1$ the nearest neighbor site to $2$ with a connecting bond $J_{12}$ and similarly
denote by $4$ the nearest neighbor site to $3$ with a connecting bond $J_{43}$. If $J_{23}$ is
much larger then the connecting bonds the spin pair $(2,3)$ acts as an effective
singlet.
It follows that the strongly correlated singlet pair
can be frozen out. Due to the virtual triplet exciations,
an effective coupling $\tilde{J}_{14}$ is generated between
the sites $1$ and $4$. These two sites  become nearest
neighbors once the singlet has been eliminated.
In a second-order perturbation calculation one obtains
\be
{\tilde J}_{14}=\kappa \frac{J_{12} J_{43}}{\Omega},\quad \kappa(S=1/2)=1/2\;.
\label{deci}
\ee
The new coupling is thus smaller than any of the original ones.
The energy scale $\Omega$
is continuously reduced upon iterating the procedure
and at the same time the probability
distribution of the couplings $P(J,\Omega)$
approaches a limiting function. In a
gapless random system $\Omega$ tends to zero at the fixed point of the transformation
and the low energy tail of the distribution is typically given by:
\be
P(J,\Omega){\rm d}J \simeq \frac{1}{z} \left(\frac{J}{\Omega}\right)^{-1+1/z}\frac{{\rm d}J}{\Omega}\;.
\label{Jz}
\ee
The dynamical exponent $z$ determines how the length scale $L$
scales with the time scale $\tau$:
\be
\tau \sim \Omega^{-1} \sim L^z\;.
\label{t_L}
\ee
In general $z$ is not a universal quantity: its value depends on the form of the
original disorder.
However $z$ stays invariant under renormalization\cite{ijl01}. Therefore
one can deduce its value from the renormalized distribution in Eq.(\ref{Jz}).
Varying the parameters of the initial
distribution one can reach a situation
where the width of the distribution in Eq.(\ref{Jz})
grows without limits, i.e. $z$ formally tends to infinity. In this case,
according to exact results on
the random AF spin-1/2 chain\cite{fisherxx}, one should formally replace $z$ in Eq.(\ref{Jz}) by
$-\ln \Omega$, so that the scaling relation in Eq.(\ref{t_L})
takes the form:
\be
\ln t_r \sim L^{\psi},\quad \psi=1/2\;.
\label{lnt_L}
\ee
This type of fixed point, where the ratio of any two neighboring bonds typically
tends to zero or infinity, is called an infinite
randomness fixed point (IRFP). It has been conjectured that the MDH
renormalization group transformation~(\ref{deci}) leads to
{\sl exact results} regarding the singular properties
of the transformation, namely the value of $\psi$
in Eq.~(\ref{lnt_L}) is exact~\cite{fisherxx,2drg}.
\begin{figure}[h]
\epsfxsize=8truecm
\begin{center}
\mbox{\epsfbox{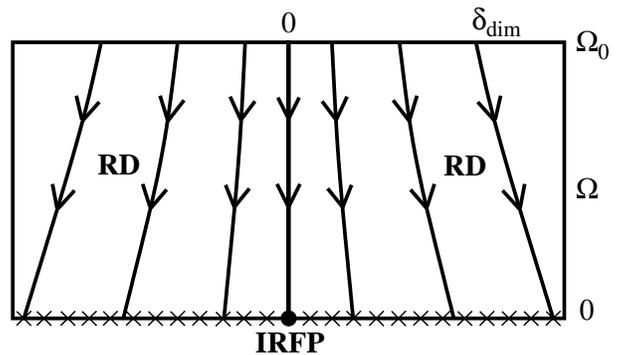}}
\end{center}
\caption{\label{fig4} Schematic RG phase diagram of the random dimerized s=1/2 chain as a
function of the quantum control parameter
$\delta_{dim}$ and the energy scale $\Omega$.
Along the RG trajectories the dynamical exponent
$z(\delta_{dim})$ is constant. The
non-dimerized model with $\delta_{dim}=0$ is attracted by the IRFP
with $1/z=0$. With
decreasing the energy scale $\Omega$,
disorder in the system is increasing.}
\end{figure}

For the random AF spin-1/2 chain, according to exact results\cite{fisherxx} any amount of
disorder is sufficient to drive the system into the IRFP. Similarly, the gapped dimer phase
will turn into a gapless random dimer phase for any amount of disorder, where the
dimerization parameter is defined as
\be
\delta_{dim}=\left[\ln J_{odd}\right]_{\rm av}-\left[\ln J_{even}\right]_{\rm av}\;
\label{delta_dim}
\ee
in terms of the couplings $J_{odd}$ and $J_{even}$ at odd and even sites,
respectively.
The random dimer phase is a quantum version of the Griffiths-phase, which has been
originally introduced for classical disordered systems\cite{griffiths}.
The schematic RG-flow diagram of the random dimerized chain is drawn in
Fig.~\ref{fig4}.
\begin{figure}[h]
\epsfxsize=8truecm
\begin{center}
\mbox{\epsfbox{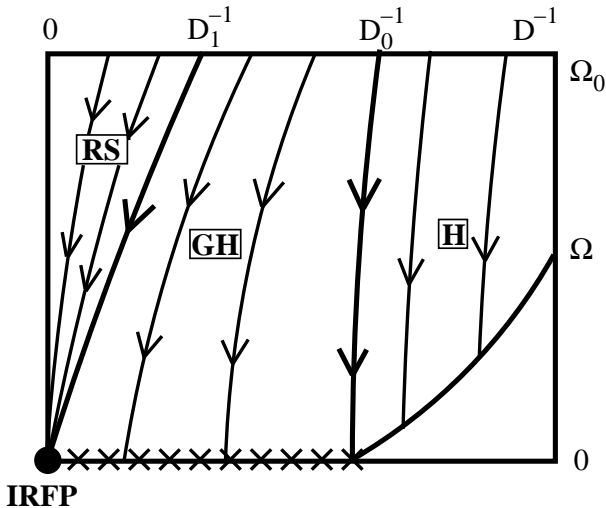}}
\end{center}
\caption{\label{fig5} Schematic RG phase diagram of the random AF
spin-1 chain, as a
function of the disorder strength $D$ and the energy scale $\Omega$.
For weak disorder, $D<D_0$, there is a Haldane ("H") gap in the spectrum. For
intermediated disorder, $D_0<D<D_1$ the system is in the gapless Haldane ("GH")
phase with a varying dynamical
exponent $z(D)$. For strong enough disorder, $D>D_1$ the system is in the random singlet
phase and scales into the IRFP.}
\end{figure}

The MDH method has also been used to study the singular properties of the random AF spin-1
chain. Here we first note that the spectrum of the pure system has
a Haldane gap, which is stable against weak randomness. Consequently the MDH renormalization,
which is by definition a strong disorder approach,
becomes valid if the initial disorder is increased over a limit, say $D_0$.
Our second remark concerns the ``projecting onto the lowest level'' procedure\cite{boechat}
when after decimating out a strongly coupled singlet the generated new coupling is in the
form of Eq.(\ref{deci}), however with a constant of $\kappa(S=1)=4/3$. Consequently at the
initial RG steps the energy scale could behave non-monotonically, so that an IRFP behavior
is expected only for strong enough initial disorder.
To cure this problem a modified
RG scheme was proposed\cite{hyman,monthus}, which is based on the principle of
``projecting out the highest level''. By this method an effective Hamiltonian with spin-1
and spin-1/2 degrees of freedom has been introduced, where between the spins both AF
and F couplings could be present but their distribution
should respect some constraints.
The RG analysis of this effective model leads to two
different types of strongly disordered
phases, provided the disorder of the original
distribution exceeds the limiting value of $D_0$.
For an intermediate range of disorder,
so that $D_0<D<D_1$ the system scales into a quantum Griffiths-phase,
the so-called
gapless Haldane phase, where $z=z(D)$ is a monotonously increasing
function of disorder and $1/z(D)>0$. When the strength of disorder exceeds a
second limit, say $D>D_1>D_0$,
the dynamical exponent becomes infinite and the singular behavior of the system is
controlled by the IRFP (see Fig~\ref{fig5}).

Till now there are no numerical estimates about the limiting disorder strengths, $D_0$
and $D_1$. For the uniform distribution, which  corresponds to $D=1$ in
Eq.(\ref{pow_dist}), the system is in the gapless Haldane phase with $z\approx 1.5$\cite{hida}. 

Finally, we mention the work by Westerberg et al.\cite{westerberg}
about renormalization
of spin-1/2 Heisenberg chains with mixed F and AF couplings.
In this problem, due to
the presence of strong F bonds under renormalization spin clusters with arbitrary large
effective moment $S^{eff}$ are generated, such that $S^{eff} \sim \Delta^{-\omega}$,
where $\Delta$ is the largest local gap in the system and $\Delta \to 0$ at the fixed
point. Singularities of different physical quantities are related to the scaling
exponent $\omega$. 

\section{Renormalization of AF spin ladders}

\begin{figure}[h]
\epsfxsize=8truecm
\begin{center}
\mbox{\epsfbox{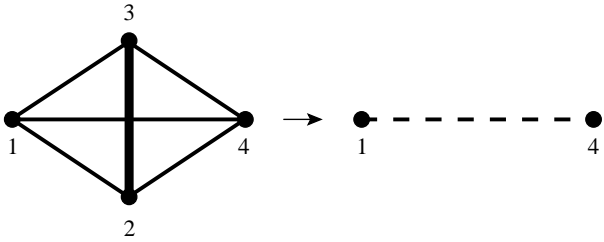}}
\end{center}
\caption{\label{fig6} Singlet formation and decimation in the ladder geometry.}
\end{figure}

With a ladder geometry,
spins are more interconnected than in a chain, which leads to a
modification of the decimation procedure described in the previous section.
As shown
in Fig.~\ref{fig6}
both spins of a strongly coupled pair, say $(2,3)$, are generally connected to
the nearest neighbor spins, denoted by $1$ and $4$. After decimating out the singlet pair the
new, effective coupling between $1$ and $4$ is of the form:
\be
{\tilde J}^{eff}_{14}=\kappa \frac{(J_{12}-J_{13})(J_{43}-J_{42})}{\Omega},\quad
\kappa(S=1/2)=1/2\;,
\label{deci_sing}
\ee
which should replace Eq.(\ref{deci}) obtained in the chain topology,
i.e. with $J_{13}=J_{42}=0$. With the rule in Eq.(\ref{deci_sing})
F couplings are also generated.
As a consequence, the renormalized Hamiltonian contains both AF and F bonds.
When at some step of the renormalization an F bond becomes the strongest one, it will
lead to the formation of an effective spin-1 
cluster. In further RG steps the system renormalizes into a set of
effective
spin clusters having different moments and connected by both
AF and F bonds.
The detailed renormalization rules have already been
given in Ref.\cite{Melin-RG}.

Due to the ladder topology and the complicated renormalization rules the RG equations
can not be treated analytically and one resorts to numerical implementations of the
renormalization procedure. We note that a variant of the MDH renormalization has been
successfully applied numerically for the two-dimensional random transverse-field Ising
model (RTIM)\cite{2drg,lkir} (also for double chains of the RTIM\cite{lkir}).
An IRFP has been obtained both for the 2D RFIM\cite{2drg,lkir}
and the double chain RFIM\cite{lkir}.

In practice  we use a finite size version of the
MDH renormalization, as for the RTIM in Ref\onlinecite{lkir}. In this method
we start with a finite ladder of $L$ sites and with periodic boundary conditions and
perform the decimation procedure
until one spin pair with a first gap $\Delta$
remains in the system.
Since $\Delta$ plays the role of the energy scale at length scale $L$,
$\Delta$ and $L$ should be related by the relation~(\ref{t_L})
involving the dynamical exponent $z$.
Performing the
above decimation for different samples the probability distribution of $\Delta$
in the small $\Delta$ limit is described by the form in Eq.(\ref{Jz}), where the energy
scale $\Omega $ is replaced by $L^{-z}$.

The IRFP is signalled by a diverging $z$, or more precisely the $P_L(\Delta) {\rm d} \Delta$
distributions have strong $L$ dependence, so that the appropriate scaling combination is
\be
\ln\left( L^{\psi} P_L(\Delta)\right) \simeq
f\left(L^{-\psi} \ln \Delta\right)\;,
\label{IRFP}
\ee
which can be obtained from Eq.(\ref{Jz}) by formally setting
$z \simeq -\ln \Delta \sim L^{\psi}$. 

In the actual calculations we have considered several 100.000
realizations of random ladders
of length up to $L=512$. Then, from the distribution of the gap at
the last step of the
RG iteration we have calculated the dynamical exponent, $z$.
The random couplings
were taken from the power-law distribution in Eq.(\ref{pow_dist}), where the strength of disorder is measured by the parameter
$D$. In the following we present our results
for the specific ladder models discussed in Sec. 2.

\subsection{Random conventional ladders}

\begin{figure}[thb]
\centerline{\fig{9cm}{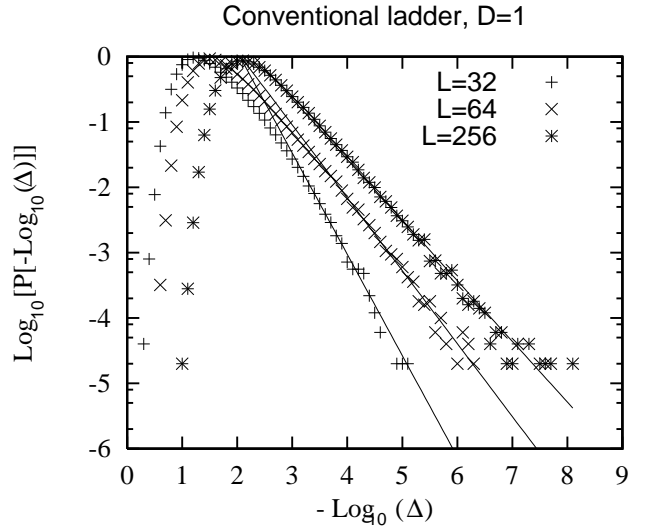}} 
\medskip
\caption{Probability distribution of the first gaps
for the conventional random ladder with a disorder
$D=1$ (see Eq.(\ref{pow_dist})) and system size
$L=32$, $L=64$, and $L=256$. For clarity, we
have not shown the data corresponding to
$L=128$.
The solid lines represent
the best fit to the form
$\log_{10}{\left[ P \left[ - \log_{10}{\Delta} \right]
\right]} = A_L - {1 \over z_L} \log_{10}{\Delta}$, with
$A_{32}=3.18$, $A_{64}=2.30$, $A_{128}=1.92$, $A_{256}=2.14$
and $z_{32}=0.65$, $z_{64}=0.90$, $z_{128}=1.06$, and
$z_{256}=1.07$. We deduce that the asymptotic value
of the dynamical exponent is $z_{\infty} \simeq 1.07$.
} 
\label{fig:PDelta}
\end{figure}

We start with the conventional ladders in Fig.~\ref{fig1}.a
where the couplings
along the
chains ($J_l^{\tau},~\tau=1,2$) and the couplings along
the rungs ($J_l^R$) are taken from the
same random distributions. In Fig.~\ref{fig:PDelta}
we show the probability distribution of the
gaps at the last step of the RG iteration calculated with the
disorder parameter  $D=1$ (see Eq.(\ref{pow_dist})).
As seen in the figure the small energy tail
of the distribution follows the functional form given by
Eq.~(\ref{Jz}) and the
dynamical exponent $z$  given by the asymptotic slope of the distributions
is finite and has only a very weak size dependence.

\begin{figure}[thb]
\centerline{\fig{9cm}{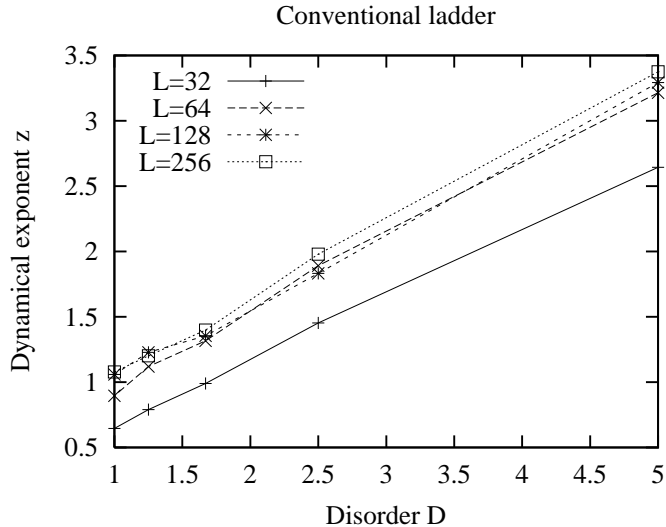}} 
\medskip
\caption{Variation of dynamical exponent
$z$ {\sl versus} disorder $D$ for
the conventional ladder with lengths $L=32$, $L=64$, $L=128$,
and $L=256$. For large system sizes and strong disorder, one has
$z_{\infty} \simeq 0.42 + 0.58 D<D$. In the region with $z_{\infty}<1$, where the
disorder is irrelevant, the system is in the gapped Haldane phase (see text).
} 
\label{fig:z-D}
\end{figure}

Repeating the calculation for other
values of $D$  we obtain a set of $D$-dependent dynamical exponents
which are represented on Fig.~\ref{fig:z-D}.
For strong disorder we obtain $z(D) < D$, which means that {\sl disorder
is reduced in the course of the renormalization}.
In the terminology of Motrunich {\sl et al.}\cite{2drg},
this system is a {\sl finite randomness} system, as opposed
to the {\sl infinite randomness} systems that will
be considered in the next subsections. For weak disorder the
dynamical exponent predicted by the approximative MDH renormalization is lowered below one
for $D<D_0 \approx 1$. Here we argue that in this region the effect of disorder is irrelevant,
so that the system is in the gapped RUS phase. Indeed, in  a pure quantum system, where scaling
in time and space is isotropic, the dynamical exponent is $z_{pure}=1$. Similarly, for disorder
induced gapless systems, where disorder in the time direction is strictly correlated, the
dynamical exponent can not be smaller, than in the pure system, so that $z_{dis} \ge z_{pure}=1$.
Consequently, if the disorder induced dynamical exponent is $z_{dis}<1$, then disorder
could only influence the correction to scaling behavior, but the system stays gapped.
In view of this remark $D_0$ can be considered as the lower limiting value of the
disorder, where the conventional finite randomness behavior ends. So that
the phase diagram of random conventional two-leg spin ladders consists of two
phases: a gapped RUS (Haldane) phase and a random
gapless Haldane phase. The latter is characterized by a finite dynamical
exponent $z(D)$ for any strong but finite initial disorder.
Consequently there is an important difference with the random AF 
spin-$1$ chain which flows into the IRFP above a finite
critical value of randomness (see Fig.~\ref{fig5}).

\subsection{Random ladders with staggered dimerization}

In this subsection we consider conventional ladders with staggered dimerization
having a dimerization parameter, $0 \le \gamma \le 1$, in Eq.(\ref{Jdimer}).
The different type of couplings in the ladder are taken from the power-law
distribution in Eq.(\ref{pow_dist}), each having the same disorder parameter, $D$,
however the range of the distribution for the different type of couplings are,
$0<J_l^R<J_{max}^R$ for the rung couplings, $0<J_l^{weak}<(1-\gamma)J_{max}$
and $0<J_l^{strong}<(1+\gamma)J_{max}$ for the weaker and stronger chain
couplings, respectively. For a fixed value of $\gamma$ and $D$ we have calculated
the finite-size dependent effective dynamical exponent, $z$, as a function of the
coupling ratio, $J_{max}^R/J_{max}$.

\begin{figure}[h]
\epsfxsize=9truecm
\begin{center}
\mbox{\epsfbox{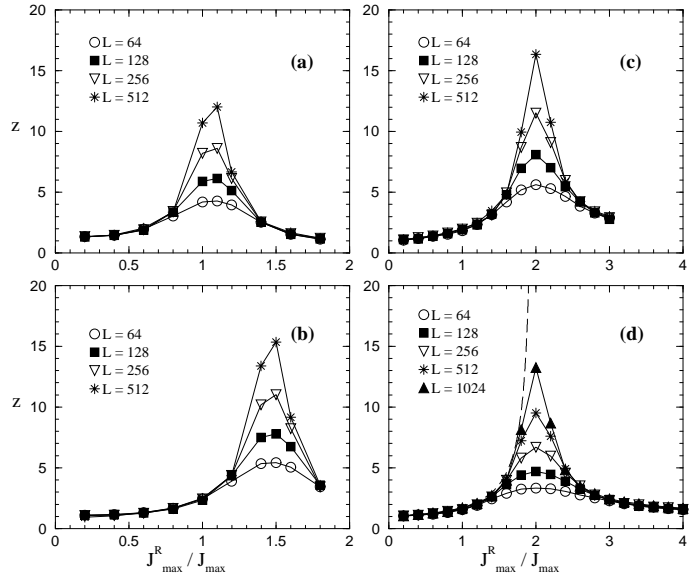}}
\end{center}
\caption{\label{fig:fig9} Finite-size estimates of the dynamical exponent of random conventional
ladders with staggered
dimerization as a function of the coupling ratio, $J^R_{max}/J_{max}$, with a disorder parameter
$D=1$ and for different dimerizations: a) $\gamma=0.5$, b) $\gamma=0.75$, c) $\gamma=1.$. In
d) a similar calculation for the random $XX$-chain is presented ($D=1,~\gamma=1$), where
the exact dynamical exponent in Eq.(\ref{z_XX}), obtained in the $L \to \infty$ limit, is shown by dashed line.}
\end{figure}

\begin{figure}[h]
\epsfxsize=8truecm
\begin{center}
\mbox{\epsfbox{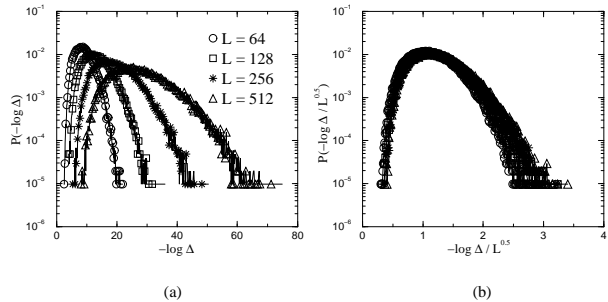}}
\end{center}
\caption{\label{fig:fig9a} a) Probability distribution of the first gap at the transition point of the
random conventional ladder with staggered dimerization, $D=1$, $\gamma=0.5$, $J^R_{max}/J_{max}=1.1$.
The distributions become broader and broader with $L$, which signals infinite randomness behavior.
b) Scaling plot in terms of the scaling combination in Eq.(\ref{lnt_L}).}
\end{figure}

As shown in Fig.~\ref{fig:fig9} the effective exponents
have the same type of qualitative behavior for different
values of the dimerization parameter, $\gamma$. In each case the curves have a maximum
at some value of the couplings, where the finite-size dependence is the strongest,
whereas more far from the maximum the convergence of the data is faster. To decide
about the possible limiting value of $z$, in particular at the maximum of the curves,
we analyze the behavior for $\gamma=1$ in Fig.~\ref{fig:fig9}c,
which is just a dimerized random
chain, the properties of which are exactly known by some extent\cite{fisherxx,ijr00}.
The random critical point of this system is situated at $J^R_{max}/J_{max}=2$, where
the critical behavior is governed by an IRFP, so that the dynamical
exponent, $z$, is formally infinity. For any other values of the couplings, $J^R_{max}/J_{max}$,
the system is in the random dimer phase, where the dynamical exponent is finite and
coupling dependent. To see the general tendency of finite-size convergence of the
$z$ exponent around the critical point we have repeated the calculation at the
$D=1,~\gamma=1$ case for the random $XX$-chain and the numerical finite-size results are compared in
Fig.~\ref{fig:fig9}d with the exact value of the dynamical exponent, as given by the solution
of the equation:
\be
\frac{J_{max}^R}{J_{max}}=2 \left( \frac{D^2}{D^2-z^{-2}} \right)^{-z}\;,
\label{z_XX}
\ee
known from Ref.\onlinecite{ijl01,ijr00}.
As seen in Fig.~\ref{fig:fig9} the dynamical exponents of random $XX$- and Heisenberg-ladders
have very similar coupling dependence and one expects the same type of divergence at the
critical point for all values of $\gamma$. In Fig.~\ref{fig:fig9a} we illustrate the scaling behavior of
the gap at the transition point, i.e. at the maximum of the curves in Fig.~\ref{fig:fig9}a. 
\begin{figure}[h]
\epsfxsize=8truecm
\begin{center}
\mbox{\epsfbox{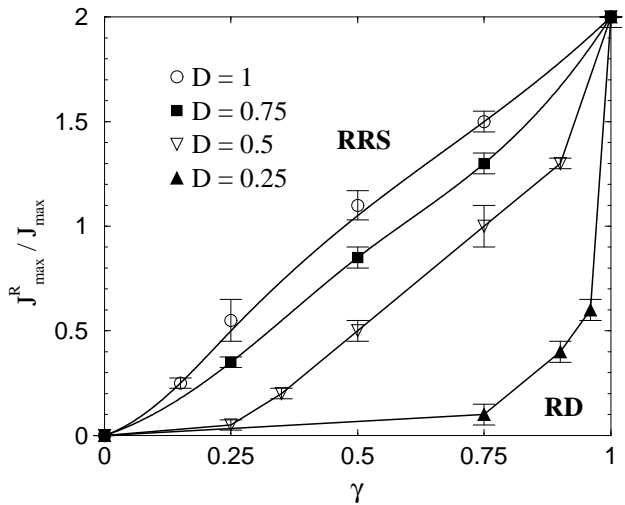}}
\end{center}
\caption{\label{fig:fig11} Phase diagram of random conventional ladders with staggered dimerization
for different disorder parameters. The random dimer (RD) phase and
the random rung singlet (RRS) phase are separated by a random critical line of infinite
randomness behavior.}
\end{figure}
The distributions
in Fig.~\ref{fig:fig9a}a become broder and broder with the size and the effective dynamical exponent is
increasing with the size without limits. An appropriate scaling collaps of the gap-distributions
has been obtained in Fig.~\ref{fig:fig9a}b, where the scaling variable in Eq.(\ref{IRFP}), with $\psi=1/2$
is used. Similar type of infinite randomness behavior is observed at other points of the critical
lines, with the same exponent $\psi=1/2$, which turned out to be universal.
Thus we conclude that the random conventional ladder with staggered
dimerization has two Griffiths-type gapless phases, the random dimer phase and
the random rung singlet phase, which are separated by a random critical line,
along which there is {\it infinite randomness} behavior. For different disorder
parameter, $D$, the position of the random critical line is modified, generally
stronger disorder is in favor of the random rung singlet phase, see
Fig.~\ref{fig:fig11}.

We note that the previously studied random conventional ladder is contained as a
special point in this phase diagram at $J^R_{max}/J_{max}=1$ and $\gamma=0$.
This point is in the random rung singlet phase for any value of $D$, thus
the dynamical exponent is finite in accordance with the previous results.

\subsection{Random zig-zag ladders}

For the zig-zag ladders the
nearest neighbor couplings $(J_l^R=J_l^Z \equiv J_l^1)$
are taken from the power-law distribution in Eq.(\ref{pow_dist}) with 
the coupling $J_l^1$ within the range $0<J_l^1<J_{max}^1$. Similarly,
the next-nearest neighbor couplings ($J_l \equiv J_l^2$)
are taken from the same
type of power-law distribution and the range of couplings is now
$0<J_l^2<J_{max}^2$. 

\begin{figure}[tb]
\centerline{\fig{9cm}{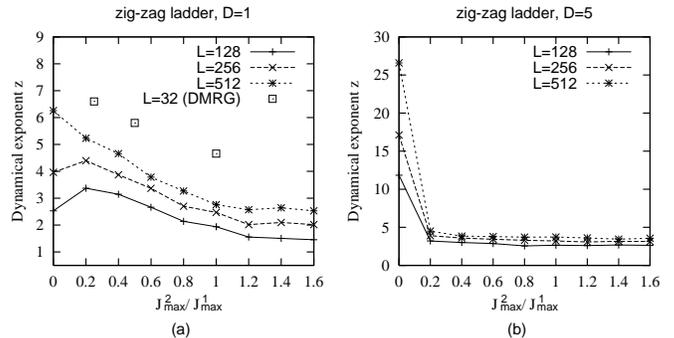}} 
\medskip
\caption{Variation of the dynamical exponent $z$
{\sl versus} $J^2_{max} / J^1_{max}$ for the zig-zag
ladder with a disorder $D=1$ (a) and $D=5$ (b).
For $D=1$, the MDH renormalization group data 
with $L=128$, $L=256$ and $L=512$ have been compared
to the DMRG calculation with $L=32$ (see Fig.~\ref{fig:DMRG}).
The lines connecting the calculated points are guides to the eyes.} 
\label{fig:zigA}
\end{figure}

The calculated dynamical exponent, $z$, as shown
in Figs.~\ref{fig:zigA}
has its maximum at $J_{max}^2/J_{max}^1=0$ and around this
point one can observe strong finite-size dependence, the range of which 
is wide, in particular for weak disorder (see Fig.~\ref{fig:zigA}).
At $J_{max}^2/J_{max}^1=0$, where the zig-zag ladder reduces to
a random AF chain, the system is in the IRFP,
thus the extrapolated value of the dynamical exponent is formally infinity.
Given the strong finite-size corrections in the numerical RG data of
the dynamical exponent\cite{remark2} in Fig.~\ref{fig:zigA},
it is difficult to decide whether the IRFP behavior
of the zig-zag ladders is extended to a finite region of the couplings
$J_{max}^2/J_{max}^1>0$ or whether this region
shrinks to a single point only. The first
scenario may be related to the existence of a gapless phase of the pure
model for $J^2 / J^1 < 0.24$.

\begin{figure}[h]
\epsfxsize=8truecm
\begin{center}
\mbox{\epsfbox{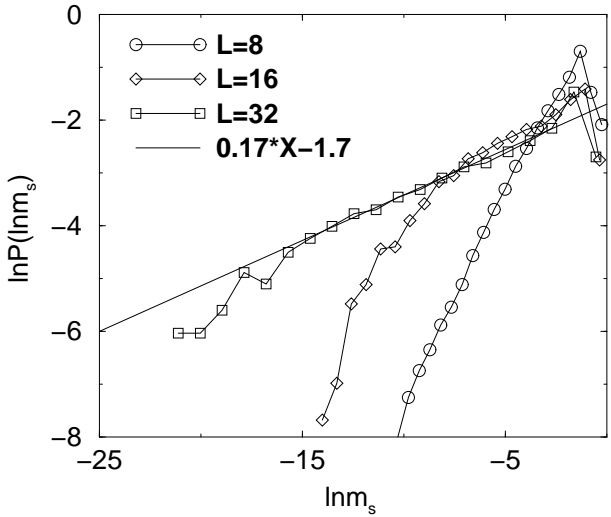}}
\end{center}
\caption{\label{fig10}\label{fig:DMRG}
Distribution of the log surface magnetization of the random
zig-zag ladder using a power-law distribution
($D=1$, $J^2_{max}/J^1_{max}=0.5$), for
different lengths of the ladder, $L$. The asymptotic slope
of the distribution, indicated
by the straight line, is the inverse of the dynamical exponent,
see Eq.(\ref{ms_distr}). }
\end{figure}
To discuss this issue we have calculated the dynamical exponent by an
independent method based on density matrix renormalization. In principle,
the dynamical exponent is related to the distribution of the first gap, $\Delta$,
in the small $\Delta$ limit, see in Eq.(\ref{Jz}) with $\Delta \to J$. However,
a precise numerical calculation of a small $\Delta$ by the DMRG method is very
difficult, therefore we used another strategy, as described in details
in Refs.\onlinecite{ijl01,cli01}. By this method
one considers the equivalent AF chain with random first- and second-neighbor
couplings (see Fig. 1d) and with fixed-free boundary conditions and
calculate the surface magnetization,
$m_s$, at the free end, which can be done very accurately by the DMRG method.
As argued in  Refs.\onlinecite{ijl01,cli01} for a random chain $m_s$ and $\Delta$ can be
considered as dual quantities, so that the distribution of the surface magnetization
is asymptotically given by
\be
P(\ln m_s ) \sim m_s^{1/z},\quad m_s \to 0.
\label{ms_distr}
\ee
Thus the dynamical exponent $z$ can be obtained from an
analysis of the small $m_s$ tail of the distribution, as illustrated in
Fig.~\ref{fig10} where the distribution function of $\ln m_s$ is given in a log-log plot for
different lengths of the ladder.
As seen in this Figure the slope of the distribution
is well defined for larger systems, from which one can obtain an accurate estimate for
the dynamical exponent, which is finite. Repeating the calculation for other values
of the coupling ratio,  $J_{max}^2/J_{max}^1$, we have obtained a set of the
dynamical exponents, which are plotted in Fig.~\ref{fig:zigA}.
These accurate DMRG data show that
the extrapolated values of the effective exponents calculated by the numerical RG-method 
are finite for any
$J_{max}^2/J_{max}^1>0$. Consequently the random zig-zag ladder
has just one IRFP at $J_{max}^2/J_{max}^1=0$, whereas
the system in the region of $J_{max}^2/J_{max}^1>0$ is in a gapless random dimer phase.
In view of the numerical results in
Fig.~\ref{fig:zigA}, where
$z_{\infty}$ seems to stay over $z_{pure}=1$, it is quite probable that the randum dimer
phase exists for any small value of the disorder.

\subsection{Random $J_1$--$J_2$ ladders}

\begin{figure}[tb]
\centerline{\fig{9cm}{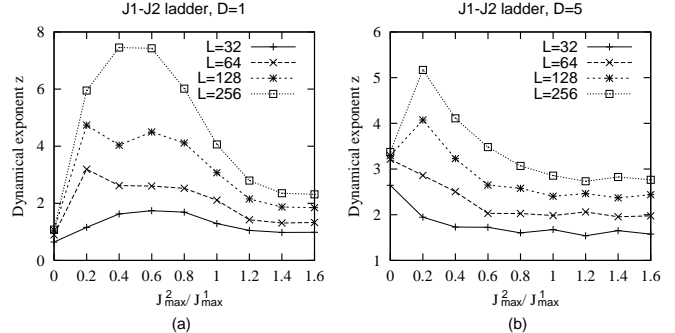}} 
\medskip
\caption{Variation of the dynamical exponent $z$
{\sl versus} $J^2_{max}/J^1_{max}$ for the random
$J_1$--$J_2$ ladder with a disorder $D=1$ (a)
and $D=5$ (b). The lines connecting the calculated
points are guides to the eyes.
Note the strong finite-size corrections in the
Griffiths phases\cite{remark2}.
} 
\label{fig:ladA}
\end{figure}

The full ladder, as represented in Fig.~\ref{fig1}.e
has three different type of couplings:
$J_l,~J_l^R$ and $J_l^D$. Here we consider a
special case of this model, when the
chain $(J_l)$ and rung $(J_l^R)$ couplings are taken from the same power-law
distribution with a disorder parameter $D$ and having a range of
$0<J_l,J_l^R<J_{max}^1$.
On the other hand the diagonal couplings are taken
from the same type of power-law distribution
and are within the interval $0<J_l^D<J_{max}^2$.
This model, having first- and second-neighbor interactions, is called a
$J_1$--$J_2$ ladder.
We have calculated the
finite-size dynamical exponents as a
function of the coupling ratio $J^2_{max}/J^1_{max}$ for
different strengths of
disorder (see Fig.~\ref{fig:ladA}).
These curves show similar qualitative behavior
as those calculated for the random conventional ladders
with staggered dimerization
in Figs.~\ref{fig:fig9}, so that we can draw similar conclusions.

The extrapolated
position of the maximum of the $z$ curves is identified as a
quantum critical point with infinite randomness behavior.
Indeed, repeating the calculation as indicated for the
dimerized ladder model in Fig.~\ref{fig:fig9a} we obtained
a scaling behavior as in Eq.(\ref{lnt_L}). with an
exponent which is compatible with $\psi=1/2$. The random quantum
critical point separates
two gapless Haldane phases, having odd and
even topological order, respectively.

\begin{figure}[tb]
\centerline{\fig{9cm}{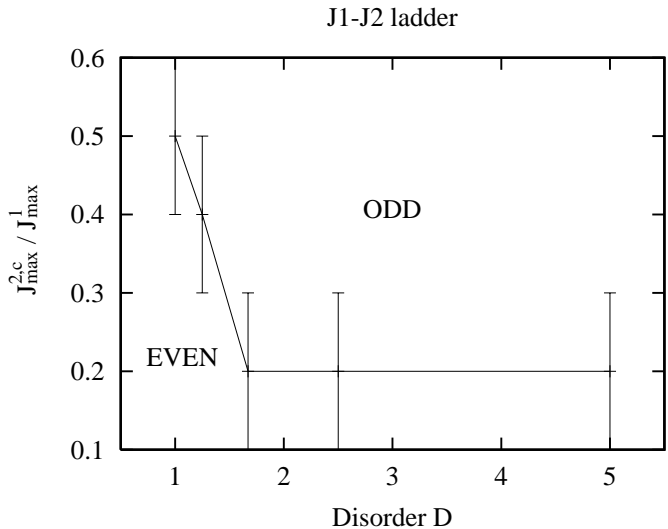}} 
\medskip
\caption{Phase diagram of the random $J_1$--$J_2$ ladder
obtained from the maximum in the variation of the
dynamical exponent {\sl versus} $J^2_{max}/J^1_{max}$
for the largest available sizes $L=256$
(see Fig.~\ref{fig:ladA} for $D=1$ and $D=5$). The two
straight lines connecting the calculated transition
points are guides to the eyes.
The whole transition line between the two phases with
even and odd topological orders, respectively, is presumably a line
of IRFPs. 
} 
\label{fig:phase-diagram}
\end{figure}

Repeating the calculation for different disorder parameters we obtain a phase
diagram shown in Fig.~\ref{fig:phase-diagram}.
In the range of disorder we used in the calculation the random
critical point is always
attracted by the IRFP, this property probably remains true for
any small value of disorder.

\section{Discussion}

In this paper different type of random AF spin ladder models
have been studied by a numerical
strong disorder RG method. In particular
we asked the questions i) how the phase
diagrams of the pure models are modified due
to quenched disorder and ii) how the
concepts observed in random AF chains, such as
infinite randomness and Griffiths-type
singularities are valid for these more complicated,
quasi-one-dimensional models.

In our numerical calculations we observed as a general
rule that for strong enough
disorder the ladder models, like the random chains,
become gapless. The dynamical
exponent of the models is generally non-universal:
$z$ depends on both the strength
of disorder and on the value of the couplings.
In models where there
is a  competition
between different types of phases, either due to
staggered dimerization or due
to frustration, such as in the $J_1$--$J_2$
model, at the phase boundary the critical
behavior of the random model is generally controlled by an infinite randomness
fixed point, at least for strong enough disorder. The low-energy properties of
the systems in this IRFP are asymptotically exactly known from analytical
calculations in random AF spin chains\cite{fisherxx,ijr00}. Thus the general
phase diagram consists of Griffiths-type phases with different topological
order separated by a random critical point of the IRFP type. The zig-zag
ladder is an exception, where there is just one Griffiths phase and
the random critical point is located at its boundary.

Next we turn to discuss about possible cross-over effects when the strength
of disorder is varied. These problems can not be directly studied by the simple
strong disorder RG method, however, from arguments considering the sign of $z_{dis}-z_{pure}$
and from analogous investigations on quantum spin
chains\cite{hyman,monthus,cli01} we can suggest the following picture. Originally
gapped phases could stay gapped for weak disorder and become gapless only if the strength
of disorder exceed some finite limiting value, as seen for the random conventional
ladder. However, for frustrated ladders, such as the zig-zag and the $J_1-J_2$ ladders,
any small amount of disorder seems to bring the system into a random gapless phase.
At a phase boundary, such as in the staggered dimerized ladder and the
$J_1$--$J_2$ model, the random critical behavior is of the IRFP type, probably
for any small amount of disorder.

At this point we comment on the similarity of the low-energy behavior of spin
chains with $2 S=odd$ ($2 S=even$) spins and that of spin-1/2 ladders with $n=odd$ ($n=even$) legs.
If the pure systems are gapless, i.e. $2S=n=odd$, strong enough disorder is expected
to bring both systems into the IRFP. For $2S=n=odd \ge 3$, there is a limiting
disorder strength, $D_c(n)$, below which the system is described by a conventional
random fixed point with $z<\infty$. On the other hand for $2S=n=even$ we have only a
partial analogy: for weak disorder both systems are gapped, which turn into a gapless
Griffiths-type phase for stroger disorder. While the random ladder stays in this conventional
random phase for any strength of the disorder the random spin chain will turn into a
IRFP behavior at some finite limiting randomness. This type of infinite randomnass behavior
can, however, be seen for frustrated even-leg $J_1-J_2$ ladders at the transition point.
We can thus conclude that random ladders with even and odd number
of legs belong to different universality classes.

Finally, we comment on random square lattice antiferromagnets, which can be obtained
in the limit when the number of legs, $n$, goes to infinity. By increasing $n$
the value of the limiting strength, $D_c(n)$, is expected to increase, too,
both for $n=even$ (for frustrated ladders) and $n=odd$. In the limit, $n \to \infty$,
$D_c(n)$ very probably tends to infinity, so that
the critical behavior of that system is described by a conventional random fixed point.
Work is in progress to verify this scenario and to obtain a general physical picture
about the low-energy properties of random two-dimensional antiferromagnets\cite{progress}.

Acknowledgment: F.I. is grateful to
J.-C. Angl\`es d'Auriac and G. F\'ath
for useful discussions. This work has been
supported by a German-Hungarian exchange program (DAAD-M\"OB), by the German Research
Foundation DFG, by the Hungarian
National Research Fund under grant No OTKA
TO23642, TO25139, TO34183, MO28418 and M36803, by the Ministry of Education under grant
No. FKFP 87/2001 and by the Centre of Excellence ICA1-CT-2000-70029.
Numerical calculations are partially performed on the Cray-T3E at Forschungszentrum J\"ulich.

\end{multicols}
\widetext


\begin{references}
\bibitem{remark1}
	The notations about "odd" and "even" string topological order are related to
	the positions of the composite spin-1 operators with an eigenvalue of
	$S^z=\pm 1$, see in Ref. \onlinecite{kim}.
\bibitem{remark2}
	The strong finite-size dependence of the RG-data indicates that several decimation
	steps are needed before the asymptotic region of the RG-flow is reached. This type
	of slow convergence for the zig-zag and the $J_1-J_2$ ladders is probably connected
	with the effect of frustration, which is not there for the conventional ladder, with
	and without dimerization.
\end{references}
\end{document}